\documentstyle[12pt]{article}
\def\Journal#1#2#3#4{{#1} {\bf #2}, #3 (#4)}

\def\NPB{{\em Nucl. Phys.} B}

\def\PLB{{\em Phys. Lett.}  B}
\def\PRL{\em Phys. Rev. Lett.}
\def\PRD{{\em Phys. Rev.} D}
\def\ZPC{{\em Z. Phys.} C}

\def\PZh{{\em Pisma Zh. Eksp. Teor. Fiz.}}

\def\IJMPA{{\em Int. J. Mod. Phys.} A}
\def\MPLA{{\em Mod. Phys. Lett.} A}
\def\UFN{{\em Usp. Fiz. Nauk.}}
\def\be{\begin{equation}}
\def\ee{\end{equation}}
\def\bea{\begin{eqnarray}}
\def\eea{\end{eqnarray}}
\begin{document}
\begin{flushright}
Preprint INR-0932/96\\
September 1996
\end{flushright}
\begin{center}
{\bf Deep-inelastic scattering data and the value of $\alpha_s$}
\end{center}
\begin{center}
{ A.~L.~ KATAEV }
\end{center}
\begin{center}
{\em Institute for Nuclear Research of the Academy
of Sciences of Russia,\\ 117312
Moscow, Russia}
\end{center}
\begin{center}
\vskip .5cm
{\bf Abstract}\\

{The brief overview of the definite determinations
of the QCD coupling constant $\alpha_s$ from the characteristics
of deep-inelastic scattering processes is given.}
\end{center}
\vskip 5cm
Invited talk at the DPF-96 Meeting of APS, Minneapolis, MN, USA,
August 10-15, 1996
\newpage
{\bf 1.}~~~Among the classical ways of ``measuring $\alpha_s$-value
is the analysis of the experimental data for the characteristics of
the deep-inelastic scattering (DIS) processes. Up to recently the
average value of $\alpha_s$, extracted from various DIS data
for the structure functions of DIS was
$\alpha_s(M_Z)=0.112\pm0.002(exp)\pm 0.007(theory)$ [1].
The NLO analysis of the most
precise experimental data for $\nu N$ DIS structure functions,
obtained by the CCFR collaboration at the Fermilab Tevatron,
gave small value of $\alpha_s(M_Z)$, namely $\alpha_s(M_Z)=0.111\pm
0.004$ [2], which was included in the
above mentioned comparative discussion
of Ref.[1].
These results are over $2\sigma$ lower than the central value
of $\alpha_s(M_Z)$, extracted at the next-to-next-to-leading order
(NNLO) of perturbative QCD from the LEP data for the hadronic decay
width of the $Z^0$-boson. Indeed, one of the most detailed analysis
of the LEP data gave $\alpha_s(M_Z)=0.120\pm0.007(exp)\pm 0.002(EW)
\pm 0.002(QCD)^{+0.004}_{-0.003}(m_t,M_H)$ [3], which is
in agreement with other determinations of $\alpha_s$ from
$Z^0$-boson decay width with fixing the value of the
top-quark mass (for the most
recent review see Ref.[4]).  However, the ``small'' DIS results are
in qualitative agreement with the small values of the parameter
$\Lambda_{\overline{MS}}^{(3)}$, which were advocated some time ago
by the QCD sum rules community\cite{ShVZ,EKV} (for the definite
application see the work of Ref.[7], which has some physical outcomes
similar to the ones of
Ref.[8], obtained with the help of the finite energy sum rules
approach [9]) and
in particular with the small value $\alpha_s(M_Z)\approx 0.109$,
extracted recently from the QCD sum rules analysis of the production
cross-section of the bottomium states in
$e^+e^-$-annihilation\cite{Voloshin}.

Other important characteristics of the DIS are the Gross-Llewellyn
Smith  sum rule $GLS(Q^2)=(1/2)\int_0^1F_3^{\nu p+\overline{\nu}p}(x,Q^2)dx$
and the polarized Bjorken sum rule $Bjp(Q^2)=
\int_0^1 g_1^{ep-en}(x,Q^2)dx$. The physical advantage of these
quantities is that besides higher order perturbative QCD corrections,
calculated at the NLO in Ref.[11], NNLO in Ref.[12] and estimated at
the $N^3LO$ level in Refs.[13,14], the non-perturbative higher-twist
contributions to these sum rules are also known and
are under better theoretical
control, than in the case of the structure functions themselves.
Indeed, the definite estimates of the twist-4 contributions to the
$GLS$  and $Bjp$ sum rules were first obtained with the help
of the 3-point functions QCD sum rules formalism in Refs.[15],[16]
respectively. As was shown in Ref.[17], the information about
the values of these non-perturbative effects is very important in the
process of the extraction of the value of $\alpha_s(M_Z)$ from the
experimental result for the $GLS$ sum rule at low energies. Indeed, using the
published experimental result of the CCFR collaboration
$GLS(Q^2=3~GeV^2)=2.50 \pm 0.018 (stat)\pm 0.078(syst)$\cite{CCFR1},
the authors of Ref.[17] obtained the following NLO and NNLO values of
$\alpha_s(M_Z)$:
$\alpha_s(M_Z)_{NLO}=0.116 \pm 0.001(stat)\pm
0.005(syst)\pm 0.003(twist)\pm 0.002(scheme)$;
$\alpha_s(M_Z)_{NNLO}=0.115 \pm 0.001(stat)\pm
0.005(syst)\pm 0.003(twist)\pm 0.0005(scheme)$; where the
scheme-dependence was estimated by comparing the outcomes of
applications of the $\overline{MS}$-scheme and the PMS vs the
effective charges approaches
(it is also of interest to think about the possibility of the
applications of BLM in the similar analysis).  The theoretical
results
of Ref.[17] are  revealing the typical features of the
behavior of the theoretical uncertainties: at the NNLO-level the
uncertainty in the values of the higher-twist contributions is
starting to play the dominant role, since the scheme-dependence
uncertainties are drastically reduced at the NNLO order.  The similar
conclusions  were also recently formulated in the process of the
analysis of the existing  experimental data
for the $Bjp$ sum rule\cite{E1,E2}.
In the works of Ref.[19,20] the values of $\alpha_s(M_Z)$, which are
very closed to the ones of Ref.[17], were obtained, namely
$\alpha_s(M_Z)=0.116^{+0.003}_{-0.005} (exp) \pm 0.003
(theory)$~\cite{E1} and
$\alpha_s(M_Z)=0.118^{+0.004}_{-0.007} (exp) \pm 0.002 
(th)$~\cite{E2}, where the uncertainty in the higher-twist 
contributions are playing the dominant role in the  theoretical 
errors.  We should warn, however, that the independent study of the 
problem of the combined description of the available experimental 
data for the $Bjp$ sum rule resulted in the more careful point of 
view, that at the existing experimental accuracy it is impossible to 
choose from the data the true value of 
$\Lambda_{\overline{MS}}^{(3)}$ (namely either 
$\Lambda_{\overline{MS}}^{(3)}\approx 200~MeV$ or 
$\Lambda_{\overline{MS}}^{(3)}\approx 400~MeV$) and of the twist-4
corrections [21].

{\bf 2.}~~~Let us now return to the discussion of the current
situation with the analysis of the experimental data of the CCFR
collaboration. In Ref.[22] using the
Jacobi-polynomial expansion method\cite{Jacobi} and the available
data of the CCFR Collaboration \cite{CCFRold} the result for
$GLS(Q^2=3~GeV^2)$, obtained in Ref.[18], was confirmed.
Moreover, using the definite extrapolation procedure of the concrete
experimental data of Ref.[2], the authors of Ref.[22] also estimated
the $Q^2$-dependence of the $GLS$ sum rule. The work of Ref.[22]
initiated the analysis of the possibility of the direct experimental
determination of the $Q^2$-dependence of the $GLS$ sum rule from the
experimental results of Ref.[2], which was made by the CCFR
collaboration in the work of Ref.[24] and resulted in rather low average value
of $\alpha_s(M_Z)$, namely $\alpha_s(M_Z)=0.110^{+0.006}_{-0.009}$.

At the next stage  the  NNLO analysis of the
CCFR data of Ref.[2] was done\cite{KKPS}. In the process of this
work,  the information about the available at present
NNLO contributions to the coefficient function of the
Mellin moments of $xF_3$ structure function \cite{VZ} and the
definite non-singlet (NS) anomalous dimensions\cite{LRV} was taken
into account. The
obtained NNLO results of Ref.[25] read:
\begin{eqnarray}
xF_3~: \alpha_s(M_Z)_{NNLO}=0.109\pm 0.003(stat)\pm 0.005(syst)\pm
0.003(th)
\nonumber \\
xF_3+F_2~: \alpha_s(M_Z)_{NNLO}=0.111\pm 0.002(stat)\pm 0.003(syst)\pm
0.003(th)
\end{eqnarray}
At the NLO level all results of the fits of Ref.[25] were in very good
agreement with the results, obtained by the CCFR-collaboration in
Ref.[2], and with the quoted above $GLS$ sum rule value of
$\alpha_s(M_Z)$, given in Ref.[24]. The further
application of the CCFR data of Ref.[2] allowed the authors of Ref.[28]
to use the proposed in this work spline
$\overline{MS}$-scheme in the concrete fits.
It should be stressed, that the spline $\overline{MS}$-scheme is
formulated to estimate the theoretical uncertainties of the
application of the standard procedure of Ref.[29], which assumes that
the behavior of the coupling constant $\alpha_s$ above and beyond
the production of the $b$-quark is matched directly in the
$\overline{MS}$-scheme at the point $M=m_b$. In fact the estimated
contribution $\Delta\alpha_s(M_Z)=+ 1\%$\cite{SSM} turned out to be
in good agreement with the one $\Delta\alpha_s(M_Z)=\pm 1.5\%$,
obtained in Ref.[1] by varying the matching point within the {\it ad
hoc} chosen interval $M=(0.75-2.5)m_b$.

However, quite unexpectedly, the CCFR collaboration announced
recently in their talks\cite{CCFRT}, that mainly due to the
corrections in the energy calibration of the neutrino beam their old
data (and thus the results of Ref.[2]) changed
{\underline{drastically}} and that the new NLO results are now
{\underline{essentially}} {\underline{larger}}
\begin{eqnarray}
xF_3~:\alpha_s(M_Z)_{NLO}=0.118\pm 0.0025(stat)\pm
0.0055(syst)\pm 0.004(th) \nonumber \\
xF_3+F_2~:
\alpha_s(M_Z)_{NLO}=0.119\pm 0.0015(stat)\pm 0.0035(syst)\pm
0.004(th)
\end{eqnarray}
and are in better agreement with the LEP average value
$\alpha_s(M_Z)\approx 0.120$. This announcement is disfavouring not
only the outcomes of the CCFR analysis of Ref.[2], but the results of
the determination of the experimental values of the GLS sum
rules\cite{CCFR1,CCFRN} from the data of the CCFR collaboration of
Ref.[2]. However, the main physical conclusions obtained in the
works of Refs.[17,22,25,28] in the process of the detailed study of
the CCFR data of Refs.[2,18] do not loose their methodological
importance.

To our point of view, it is rather important to publish the modified
CCFR data (which are still even not distributed) and to check
carefully the results of Eq.(2) both at the NLO and NNLO with the
help of the methods, used in Ref.[25] in the process of the NNLO
analysis of the previous CCFR data of Ref.[2]. It should be stressed
that the results of Eq.(2) are lying higher than the previous 
determinations of $\alpha_s(M_Z)$ from the NLO fits of the
combined BCDMS-SLAC data for the structure function $F_2$ of the $\mu
N$ DIS, namely $\alpha_s(M_Z)=0.113\pm 0.003 (exp) \pm
0.004(theory)$\cite{VM}. Notice, that the NS
NNLO fits of Ref.[32] of the BCDMS-SLAC data are demonstrating, that
the NNLO perturbative QCD corrections have the tendency to decrease
this value.

It shoul be also stressed, that
that the results of Eq.(2) are excluding the ``small''
values of $\Lambda_{\overline{MS}}^{(3)}\approx 200-250~MeV$, which
are usually
used in the process of different applications of the QCD sum
rules method of Ref.[5]. In order to understand whether the main
physical predictions of the QCD sum rules method of Ref.[5] are
really able to challenge the possibility that the values 
of $\Lambda_{\overline{MS}}^{(3)}$  
can be ``large'' (namely 
 $\Lambda_{\overline{MS}}^{(3)}\approx 350-450~MeV$) 
it becomes now rather important to repeat the QCD
analysis of Ref.[6] using the  $e^+e^-$-annihilation low-energy
Novosibirsk data and to check the sensitivity of the QCD sum rules
predictions for the properties of different hadrons to the variations
of the values of $\Lambda_{\overline{MS}}^{(3)}$ and the gluon
condensate parameter $<\alpha_sG^2>$. To our knowledge, this work is
already in progress\cite{Shpr}.

Another important problem is related to the necessity of more 
detailed analysis of low $x$ and high $Q^2$ HERA 
data for $F_2$ structure function of $eN$ DIS. Indeed, the 
determination of the value of 
of $\alpha_s(M_Z)$ from the fits of the 1993-year data resulted 
in the value $\alpha_s(M_Z)=0.120\pm 0.005 (exp) \pm 0.009(th)$
\cite{BF}, which has large theoretical uncertainties. The work 
on the incorporation in this analysis of the 1994-year data is now 
in progress\cite{RB}. The attempt to make the independent similar 
analysis was made in Ref.[36]. It resulted in the following estimate 
$\alpha_s(M_Z)=0.113\pm 0.002(stat)\pm 0.007(syst)\pm
0.007(th)$\cite{HERA}. However, this estimate is still not the 
final result,
since the full NLO results are  not yet included in the
fits of Ref.[36]. 

As to the recent
combined fits of the data of BCDMS, NMC, CCFR, H1, ZEUS
collaborations using the sets of CTEQ4 parton distributions (see
Ref.[37]) and MRS parton distributions (see Ref.[38]), 
in view of the announcement  of the CCFR collaboration, 
that their ``old'' data are now corrected\cite{CCFRT}, they 
should be redone in future.

We also hope that   that the possible future more detailed 
Jacobi polinomial NNLO QCD analysis of the BCDMS data 
will be able to clarify what is the real place of the "small" 
values of $\alpha_s(M_Z)=0.111-0.113$,  
which are so
important in the analysis of the possibility of the
virtual manifestation of the effects of SUSY particles (see e.g.
Refs.[39] and the reviews of Ref.[40,41]).

It is also very important to perform the careful analysis of the 
existing low-energy Protvino data for $\nu N$ DIS with taking into 
account NNLO perturbative QCD effects and the high-twist corrections.

%And let us stress again that to our point of view the detailed 
%consideration of the sensitivity of the QCD sum rules predictions for 
%the characteristics of hadrons to the variation of the value of the 
%parameter $\Lambda_{\overline{MS}}^{(3)}$ is coming on the agenda.

%\newpage
%\section*{Acknowledgments}

We are grateful to V.A. Kuzmin and M.A. Shifman for useful
discussions. We also wish to thank R. Ball for the clarification 
of the current status of the QCD analysis of the HERA data.
It is the pleasure to thank the members of the
University of Minnesota for hospitality during this interesting
meeting. The participation at the
DPF-96 Meeting of APS was partly supported by the Russian Fund for
Fundamental Research, Grant N 96-02-18897. The work on this report 
is done within the framework of the Grant N 96-01-01860, supported  
by the Russian Fund for Fundamental Research.

% We can insert an appendix here and place equations so that they are
%given numbers such as Eq.~\ref{eq:app}.
%\be
%x = y.
%\label{eq:app}
%\ee
%\section*{References}

\end{document}